\documentstyle[aps,multicol,epsf]{revtex}
\begin{document}
\draft
 
\title{Surface Flows of Granular Mixtures: III. Canonical Model}
 
\vspace{1cm}

\author{Thomas Boutreux$^1$, Hern\'an A. Makse$^{1,2}$, and 
Pierre-Gilles de Gennes$^1$}

\address{$^1$Laboratoire de Physique de la Mati\`ere Condens\'ee,
URA $n^o792$ du C.N.R.S., \\ Coll\`ege de France,
11 place Marcelin Berthelot, 75231 Paris Cedex 05, France\\
$^2$ Center for Polymer Studies and Physics Dept., Boston University,
Boston,  Massachusetts 02215, USA}

\date{submitted to Eur. Phys. J-B}
\maketitle

\begin{abstract}
We present the generalization of the minimal model for surface flows 
of granular 
mixtures, proposed by Boutreux and de Gennes 
[J. Phys. I France {\bf 6}, 1295 
(1996)]. The minimal model 
was valid for grains 
differing only in their surface properties. 
The present model also takes into 
account differences in the size of the grains. We apply the model to
study segregation in two-dimensional silos of mixtures of grains
differing 
in size and/or  
surface properties. When the difference in size is small, 
the model predicts that a 
continuous segregation appears 
in the static phase during the filling of a silo.
When the difference in  size is wide, we take into account
the segregation of the grains in the rolling phase, and
 the model predicts complete 
segregation and 
stratification in agreement with experimental observations.
\end{abstract}


\vspace{1cm}

SHORTENED TITLE: Surface flows of granular mixtures

\begin{multicols}{2}

\section{Introduction}

Segregation of mixtures of grains is commonly observed in granular
materials that are poured, vibrated, or 
rotated~[\onlinecite{bagnold,review1,review2,review3,review4,review5,review6,varenna,hans0,brazilnut1,brazilnut2,brazilnut2.5,brazilnut3,brazilnut4,brazilnut5,kaka1,brazilnut8,thin1,thin2}]. 
This phenomenon has  great importance in industrial processes. 
The simplest way to observe segregation is to pour a mixture of grains of
different sizes onto a heap; 
one obtains a heap with the large grains near the bottom and
the small grains at the 
top~[\onlinecite{segregation1,segregation2,segregation3,segregation4,segregation5,segregation6}].

The mixture  can also be 
poured in a two-dimensional silo made up of two vertical
plates separated by a gap of approximately 5 mm (a granular Hele-Shaw cell), 
as studied recently by 
Makse {\it et al.}~[\onlinecite{makse1,makse2,kaka2,yan}]. 
Different forms of segregation are observed when the grains
differ
both in their size and their surface properties 
(shape, roughness, stickiness).
When the grains do not differ much in size, 
the larger or the smoother 
grains stop preferentially at the bottom of the slope, 
and the smaller or the rougher at the top of the slope. 
This segregation is limited since grains 
of both species remain present everywhere; this phenomenon is called 
{\it continuous segregation}. 

When the grains have a wide difference in size, segregation is stronger and 
appears in two different ways. When the large grains are smoother than the 
small ones, a {\it complete} segregation is observed, where all large grains 
stop near the bottom of the slope, and all small grains stop near 
 the top.  When the 
large grains are rougher than the small grains, 
a spectacular {\it stratification} is obtained, 
where the grains deposit in alternating layers of different species, parallel 
to the sand-pile surface.

Theoretical studies of surface flows of grains were triggered by the
works of Bouchaud, Cates, R. Prakash, and Edwards 
(BCRE)~[\onlinecite{bouchaud1,bouchaud2}] and Mehta and
collaborators~[\onlinecite{mehta}].
BCRE proposed a set of coupled equations to describe granular flows in
one species sand-piles. Boutreux and de Gennes (BdG)
~[\onlinecite{pgg,bdg}]
generalized the BCRE equations in order to describe granular mixtures 
composed 
of two species. They proposed a theoretical formalism, and a {\it minimal} 
model 
describing the case of grains with different surface properties but 
equal size. 
In this case, BdG found a power law behavior of 
the concentrations which explains the continuous segregation. 
Since the minimal model does not take into account the size difference 
between 
the grains, 
Boutreux has
treated 
the important case where 
the grains differ only in   size, in
a second article~[\onlinecite{epjbII,TheseTom}] of the series started by 
BdG~[\onlinecite{bdg}].

In the present article, the third and last 
one of the series, we treat the general case of the canonical model, 
by describing 
different examples where the grains differ in  size and/or  
surface properties. We first assume that the
two species do not differ much in size. We then consider that
there is no segregation inside the rolling phase, which is homogeneous in
the vertical direction. In this case, we show that the canonical model
predicts continuous segregation (in agreement with experiments)
with a power law behavior of the concentrations.

In the last part of the article, we consider
the case in which the two species have a wide difference in size. 
We argue that in this case segregation probably occurs directly inside 
the rolling phase; due to percolation of the small grains, 
the small grains fall through the gaps between the large ones.
We show that by taking  this phenomenon
into account and relaxing
the assumption that the rolling phase is homogeneous,
complete segregation occurs when the small grains are rougher
than the large grains, 
and stratification occurs when the large grains are rougher than the 
small grains, in agreement with 
experiments. In these cases, our results are consistent with theoretical 
studies 
recently published 
by Makse, Cizeau, and Stanley (MCS)~[\onlinecite{mcs,m,cms}], 
who proposed a modified version of the BdG equations. 
MCS proposed a model valid for a mixture of grains with 
 very different size.  They explained the complete segregation
phenomenon
by predicting an exponential behavior of the concentrations, and 
successfully reproduced the mechanism leading to stratification as observed 
in the experiments.

The present article is organized as follows. 
In Sec.~\ref{theory}, we review the theoretical formalism developed 
in~[\onlinecite{bouchaud1,bouchaud2,mehta,pgg,bdg}]. In
Sec.~\ref{canonical-model}, 
we present our model of binary collisions between one rolling grain 
and one grain 
at rest, and we obtain the canonical model for surface flows of 
granular mixtures. 
Section~\ref{cases} describes the application of the model to all 
possible 
cases, when the two granular species differ in size and/or in 
surface properties. 
Then in Sec.~\ref{steady}, we model the steady state 
filling of a two-dimensional 
silo by a mixture of grains, and
 we discuss the predicted segregation profiles according 
to the different composition of the  
mixture. Finally in Sec.~\ref{stratification}, 
we consider the case in which the grains have a wide difference in size, 
in order to describe stratification and complete segregation.

\begin{figure}
     \centerline { \vbox{ \hbox
     {\epsfxsize=7.cm \epsfbox{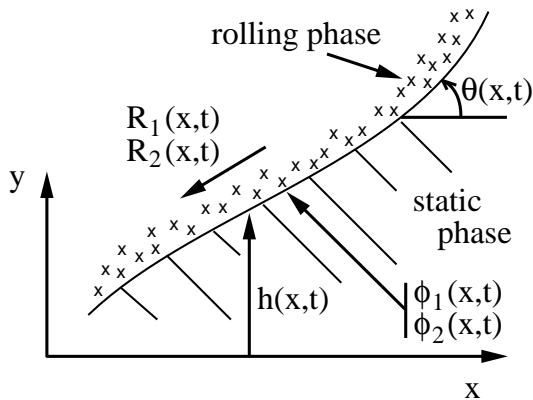} } }
     }
\vspace{1cm}
\narrowtext
\caption{Diagram showing the variables used to describe the 
granular flow of mixtures in a two-dimensional silo.} 
\label{variables}
\end{figure}

\section{Theoretical formalism}
\label{theory}

The study of two-dimensional surface flows of 
granular materials has progressed
significantly with the works of BCRE~[\onlinecite{bouchaud1,bouchaud2}] 
and Mehta {\it et al.}~[\onlinecite{mehta}].
BCRE proposed a set of variables and a set of coupled equations
to describe two-dimensional granular flows for a pure species of grains.
Recently, BdG~[\onlinecite{bdg}] generalized the BCRE formalism 
by considering a mixture of two granular species. 
In the present paper, we describe this formalism for two species.
Following Bouchaud {\it et al.}, 
we assume that there is a sharp distinction between
a static phase, where grains at rest belong to the pile, and a thin 
rolling phase where grains are not part of the pile but roll downwards 
on top of the static phase (see Fig.~\ref{variables}).
We call $\theta(x, t)$ the local slope of the interface,
and $h(x, t)$ the height of the static phase. 
For notational convenience we do not consider the difference between the
angle $\theta$ and its tangent
\begin{equation}
\theta(x,t) \simeq   -\frac{\partial h}{\partial x}.
\end{equation}

We call $\phi_{\alpha} (x, t)$
the volume fractions of the two species of grains in the static
phase just below the interface (here the index $\alpha$ denotes
the grain species, and is equal to '1' or '2'). We have
$\phi_1 + \phi_2 = 1$.
We assume that both species
have a small difference in size, so that there is no segregation 
inside the rolling phase, {\it i.e.},  the rolling phase is 
homogeneous in the vertical direction
(this assumption will be released in Sec.~\ref{stratification}).
We call $R(x, t)$ the total thickness of the rolling phase.
We also consider two ``equivalent thicknesses'' for the two species in 
the rolling phase $R_{\alpha}(x, t)$ ({\it i.e.},
 the total thickness multiplied by the
local volume fraction of the $\alpha$ grains in the 
rolling phase at position $x$).
The total thickness of the rolling phase is then equal to 
\begin{equation}
 R(x,t) \equiv R_1(x,t) + R_2(x,t).
\end{equation}

The equations describing surface flow of grains  take 
into account the fact that 
 grains in the rolling phase move 
downwards due to their weight, and 
collisions between the rolling grains and the 
static grains induce exchanges between the two phases. 
The equation that describes the exchange of grains between the two phases is
\begin{equation}
\label{dot_h}
\dot{h} = - ( \dot{R}_1|_{coll} + \dot{R}_2|_{coll} ),
\end{equation}
where the dot denotes a time derivative, and  
$\dot{R}_{\alpha}|_{coll}$  the exchange between the $\alpha$
grains in  the rolling phase with the static phase.
Equation ($\ref{dot_h}$) can also be written for the
single species 
\begin{equation}
\label{phi}
\phi_{\alpha} \dot{h} = - \dot{R}_{\alpha}|_{coll}.
\end{equation}

The evolution equation for each species in the rolling 
phase, taking into account the downhill convection of grains due to
gravity, is
\begin{equation}
\label{dot_R}
\dot{R}_{\alpha} = v \frac{\partial R_{\alpha}}{\partial x}
+ \dot{R}_{\alpha}|_{coll},
\end{equation}
where $v$ is the convection speed of the rolling grains,
which may depend on the slope  $\theta(x, t)$ and on the
grain species. However, we assume that the two species
are mixed inside the rolling phase;  the convection speed $v$
must be identical for the two species. Moreover, in practice
variations in the angle  $\theta$ are small enough, so that the
speed $v$ can then be taken as a constant.

\begin{figure}
     \centerline { \vbox{ \hbox
     {\epsfxsize=7.cm \epsfbox{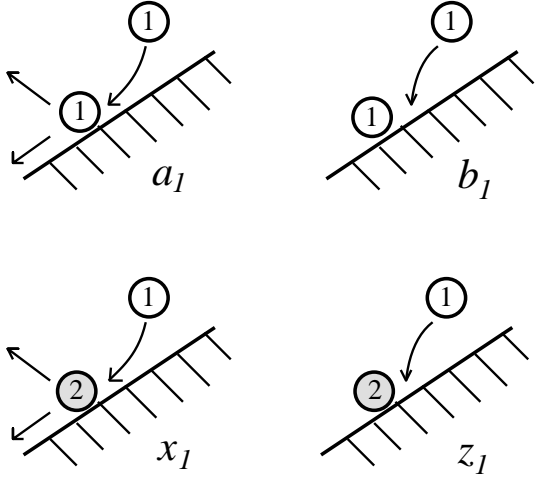} } }
     }
\vspace{1cm}
\narrowtext
\caption{Diagram showing the different type of collisions between 
a rolling grain of type 1, and a static grains of type 1 or 2. 
$a_1$: auto-amplification,  $x_1$: cross-amplification,  
$b_1$: auto-capture,  $z_1$: cross-capture.  }
 \label{four}
\end{figure}

\section{Canonical model}
\label{canonical-model}

We propose a microscopic model of the grain collisions, that will allow us to
calculate the exchange term $\dot{R}_{\alpha}|_{coll}$ 
as a function of $\theta$, $R_{\alpha}$, and $\phi_\alpha$. 
We calculate, in first order approximation, the interaction term
$\dot{R}_{\alpha}|_{coll}$ by considering  the  binary collisions
between one rolling grain and one grain at rest on top of 
the static phase. Collisions between three grains (or more) are 
less probable, and will be neglected.
We consider four types of collisions for a rolling grain of type $1$
(Fig.~\ref{four}):
\begin{itemize}
\item Auto-amplification: another type $1$ grain
starts to roll. This collision contributes with 
a term $a_1(\theta) \phi_1 R_1$ to
$\dot{R}_1|_{coll}$. This term is proportional to $R_1$ because 
all type 1 rolling grains interact with the
static phase due to the fact that 
the
rolling phase is thin.  
It is also  proportional to the concentration
of type 1 grains in the static phase,  $\phi_1$. 
The function $a_1(\theta)$
is called the {\it collision function} 
for the auto-amplification of type 1 grains. This function
 is positive, has
dimensions of a frequency, and  depends {\it a priori} 
on the angle $\theta$. The collision functions
may also depend
on the type of 
species in the static phase that are
in contact with the grain about to roll; the rougher the neighboring grains,
the less chance for  
the static grain to start rolling, and the smaller the value of $a_1$.
However, this neighbouring effect is expected to  be small
compared to the variations of $a_1$ with respect to $\theta$. In order to
simplify, we will neglect the neighbouring effect, and will 
consider that $a_1$ only
depends on $\theta$. The same approximation will be used for all the
other collision functions. 
\item Cross-amplification: a type  $2$ 
grain starts to move. This collision 
contributes with a term $x_1(\theta) \phi_2 R_1$
to $\dot{R}_2|_{coll}$. 
\item Auto-capture: the type $1$ rolling grain
is captured after a collision with a type $1$ static grain. This process 
contributes with a term $- b_1(\theta) \phi_1 R_1$ to $\dot{R}_1|_{coll}$. 
\item Cross-capture:
the type $1$ rolling grain is captured by a type 2 static grain. This process
contributes with a term $- z_1(\theta) \phi_2 R_1$ to $\dot{R}_1|_{coll}$.
This cross-capture interaction was not taken into account in the
minimal model~[\onlinecite{bdg}]. It plays an important role when the grains 
have different sizes, as will appear later. 
\end{itemize}

Collisions where the rolling grain stops and simultaneously 
a static grain starts to roll are a limiting case between cross-amplification
and cross-capture. The probability that 
these two collisions happen simultaneously is low, and therefore we 
neglect this process. 

When the colliding
rolling grain  belongs to 
the type $2$ species, four similar binary collisions occur. We call
the corresponding collision functions $a_2(\theta)$, 
$x_2(\theta)$, $b_2(\theta)$, and $z_2(\theta)$, which are all positive. 
Since increasing the slope of the pile favors rolling,  
the amplification functions $a_{\alpha}$ and $x_{\alpha}$
are increasing functions of $\theta$. Conversely,  since decreasing the slope
favors capture, the  capture functions
$b_{\alpha}$ and $z_{\alpha}$ are decreasing functions of $\theta$.

It is now possible to write the expressions of the exchange term
$\dot{R}_{\alpha}|_{coll}$ in a matrix form.
We define
the {\it collision matrix} $\hat{M}$ by~[\onlinecite{bdg}]
\begin{equation}
\left( \matrix{
\dot{R}_1|_{coll} \cr
\dot{R}_2|_{coll}  }
\right)
= \hat{M} \cdot
\left( \matrix{
R_1 \cr
R_2  }
\right).
\end{equation}
The previous microscopic description yields
\begin{equation}
\label{canonical}
\hat{M} =
\left( \matrix{
(a_1 - b_1) \phi_1 - z_1 \phi_2 & x_2 \phi_1 \cr
x_1 \phi_2 & (a_2 - b_2) \phi_2 - z_2 \phi_1
} \right).
\end{equation}
The elements $\mbox{M}_{\alpha\beta}$ are determined
by the concentrations $\phi_\alpha(x,t)$, and by the local angle $\theta(x,t)$
 via the collision functions.

Next we  calculate the expressions of the collision functions
by doing a linear approximation around the angles  of repose.
 In order to do so,
let us consider the quantity
$E_1$ defined as the total exchange from the static phase to the rolling 
phase due to
collisions originated by type $1$  rolling grains--- similar definition
can be provided for the type $2$ rolling grains, $E_2$.
A type $1$ rolling grain can interact--- via auto-amplification
or auto-capture--- with other type $1$ static
grains giving rise to a contribution
$(a_1 - b_1) \phi_1 R_1$ to $E_1$, or it can interact--- via 
cross-amplification or cross-capture--- with type $2$
static grains giving rise to a contribution 
$(x_1 - z_1) \phi_2 R_1$ to  $E_1$. 
Then $E_1$ is given by
\begin{eqnarray}
E_1 &=& [ (a_1 - b_1) \phi_1 + (x_1 - z_1) \phi_2 ] R_1  \nonumber \\
    &=& (M_{11} + M_{21}) R_1 .
\end{eqnarray}
Notice that $E_1$  contributes to $\dot{R_1}|_{coll}$ via $M_{11}$,
and also to $\dot{R_2}|_{coll}$ via $M_{21}$.

In a model of a single species made of type $1$ grains, the exchange
term would be equal to $E_1 = [a_1(\theta) - b_1(\theta)] R_1(x,t)$, 
and the angle $\theta$ for which $a_1 = b_1$
would correspond to the situation where there is no
exchange of grains between the static and the rolling phases.
We call this angle the {\it angle of repose} of the  pure type $1$ grains, 
which we denote $\theta_{11}$. 
In general, 
the repose angle $\theta_{\alpha \alpha}$ of the pure 
$\alpha$ species is the angle at which the 
auto-amplification and auto-capture functions intersect
\begin{equation}
\label{theta11}
a_{\alpha}(\theta_{\alpha \alpha}) = b_{\alpha}(\theta_{\alpha \alpha}).
\end{equation} 
Moreover, we will call the {\it cross-angle of repose} 
$\theta_{\alpha \beta}$ the angle for which the cross-amplification function 
equals the cross-capture function
\begin{equation}
\label{theta12}
x_{\alpha}(\theta_{\alpha \beta}) = z_{\alpha}(\theta_{\alpha \beta}) .
\end{equation} 
The angle $\theta_{12}$ for the type 1 grains is defined  by $x_1(\theta_{12}) 
= z_1(\theta_{12})$, and corresponds to the angle of repose
of a type $1$ rolling grains moving on top of a surface of type $2$ 
static grains [\onlinecite{mcs}].

When the surface properties of the two species are not very different, the
two pure angles of repose do not differ very much.
Moreover, when the size of the grains do not differ much, the cross-angle
of repose are also close.
 In practice, the angle $\theta$ remains close to the
angles of repose of the species and
 we can linearize the eight collision functions with respect 
to $\theta$ to get a simple expression for $E_1$
\begin{equation}
E_1 = [ \gamma_1 (\theta - \theta_{11}) \phi_1 + \gamma_2 
(\theta - \theta_{12}) \phi_2 ] R_1,
\end{equation}
where
\begin{equation}
\gamma_1 \equiv \partial_{\theta}a_1 - \partial_{\theta}b_1, \hspace{0.5cm}
\gamma_2 \equiv \partial_{\theta}x_1 - \partial_{\theta}z_1.
\end{equation}
In order to consider the simplest case, we assume that all the
derivatives of the collision functions have the same order of 
magnitude, {\it i.e.}
\begin{equation}
\partial_{\theta}a_1 \simeq  \partial_{\theta}x_1
\simeq  - \partial_{\theta}b_1 \simeq  - \partial_{\theta}z_1,
\end{equation}
so that 
$\gamma_1 \simeq \gamma_2 \equiv \gamma$, and therefore
\begin{equation}
\label{E1}
E_1 = \gamma [ \theta - \theta_1(\phi_2)] R_1,
\end{equation}
where the angle $\theta_1(\phi_2)$ is given by
\begin{equation}
\label{theta1}
\theta_1(\phi_2) \equiv \theta_{11} + ( \theta_{12} - \theta_{11}) \phi_2.
\end{equation}
The constant $\gamma$ has the dimensions of  frequency, 
and has typical value
$\gamma\sim 25$~${\rm s}^{-1}$~[\onlinecite{makse2}].
Dimensional analysis show that the order of magnitude of $\gamma$
is given by
\begin{equation}
\label{gamma}
\gamma \simeq v/d,
\end{equation}
where $d$ is the typical size of the grain.
The constant $\gamma$ represents the frequency of interaction between a 
rolling 
grain and the static phase. 
The larger the value of $\gamma$, the more frequent 
the exchange between phases.

Equation (\ref{E1}) shows that $\theta_1(\phi_2)$ is a cross-over angle;  
$E_1$ describes capture of rolling grains ($E_1 < 0$) 
when $\theta < \theta_1$, 
and amplification of grains ($E_1 > 0$) when $\theta > \theta_1$. 
The angle $\theta_1(\phi_2)$ for a mixture of grains
plays the role of the constant angle of repose $\theta_{11}$
for a pure species of grains;  
the angle $\theta_1(\phi_2)$ is called 
the {\it generalized angle of repose} 
for the type 1 species. 
When no type $2$ grains are present
on the static phase ($\phi_2 = 0$), $\theta_1(\phi_2)$ 
(given by Eq. (\ref{theta1})) is equal to
the angle of repose $\theta_{11}$ of the pure type $1$ species. When $\phi_2$
increases, the type $1$ rolling grains 
interact more frequently with type 2 static grains, and the generalized angle 
of repose $\theta_1(\phi_2)$ changes. 

The generalized angle of repose was introduced by MCS~[\onlinecite{mcs}],
who have shown that it has a key role in explaining the segregation 
as well as the stratification of granular mixtures.
MCS defined the generalized angle of repose $\theta_{\alpha}(\phi_{\beta})$
of species $\alpha$ as a linear function of the 
volume fraction $\phi_{\beta}$
and proposed the expression $(\ref{theta1})$. In the present paper we 
show that this expression can be derived, by using the binary collision model
and a linear development approximation. When it is possible, notations of 
reference~[\onlinecite{mcs}] are used in the present paper.

Making similar approximations, we get a simplified expression for the
exchange $E_2$ due to collisions with type $2$ rolling grains
\begin{equation}
E_2 = \gamma [ \theta - \theta_2(\phi_2)] R_2,
\end{equation}
where
\begin{equation}
\theta_2(\phi_1) \equiv \theta_{22} + ( \theta_{21} - \theta_{22}) \phi_1
\end{equation}
is the generalized angle of repose for the type $2$ species.

The generalized angles of repose
quantify the degree of interaction of the rolling species with the pile.
If the species $\alpha$ have a smaller generalized angle of repose for 
any value
of $\phi_\alpha$, then, when $R_1 = R_2$, the $\alpha$ rolling grains are less 
captured (or amplify more grains in the bulk), than the other species. 
In order to quantify this behavior, we define the difference
\begin{equation}
\psi_{12} \equiv \theta_1(\phi_2) - \theta_2(\phi_1).
\end{equation}
In the general case, $\psi_{12}$ is a function of $\phi_{\alpha}$. 
However in order 
to simplify, in the following we will consider particular 
situations where $\psi_{12}$ 
is a constant that does not depend on $\phi_{\alpha}$
({\it i.e.} $\theta_{12} - \theta_{11} = \theta_{22} - \theta_{21}$).

It is now possible to write a simpler expression for the 
collision matrix (\ref{canonical}).
Previous calculations for $E_{\alpha}$ yield
\begin{equation}
\label{M}
\hat{M} =
\left( \matrix{
\gamma [\theta - \theta_1(\phi_2)] - x_1(\theta) \phi_2
& x_2(\theta) \phi_1 \cr
x_1(\theta) \phi_2  
& \gamma [\theta - \theta_2(\phi_1)] - x_2(\theta) \phi_1  }
\right) , 
\end{equation}
where the cross-amplification functions can be written in the following way
\begin{eqnarray}
\label{xi}
x_1(\theta) = \frac{\gamma}{2} (\theta - \theta_{11}) + x_0, \hspace{0.5cm}
x_2(\theta) = x_1(\theta) - \Delta x_{12}.
\end{eqnarray}
Here $x_0$ and $\Delta x_{12}$ are two constants of the model.
We call Eqs. (\ref{M}) and (\ref{xi}) the canonical form 
of the collision matrix. 
This canonical form is general, and valid for any mixture of two granular 
species. 
In practice, the grains differ in  size and in surface properties. 
Let us now see how the canonical form of the collision matrix conveys these 
differences.

\begin{figure}
     \centerline { \vbox{  \hbox{ \epsfxsize=7.cm
     \epsfbox{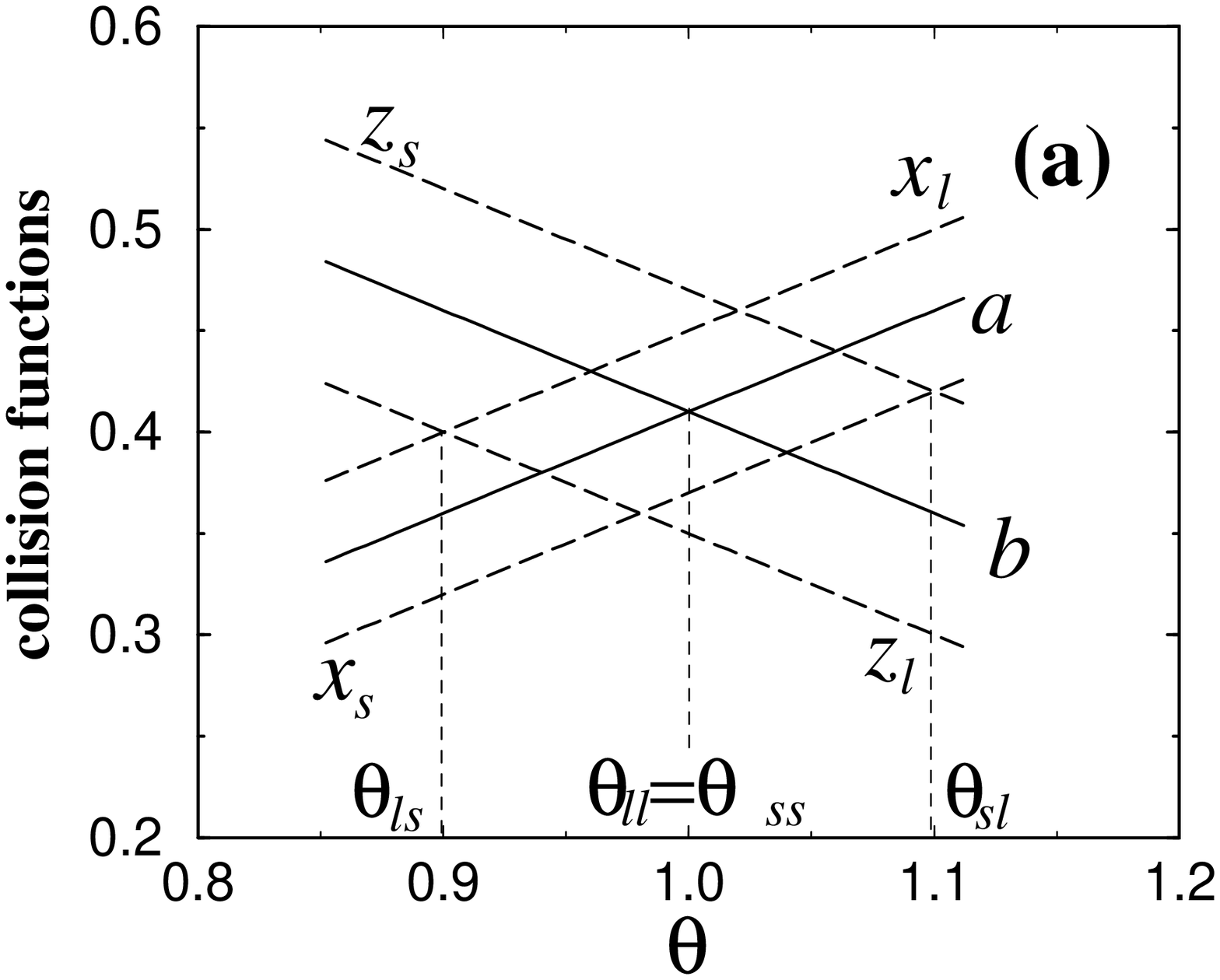} } }
     }
     
     \centerline{ \vbox{ \hbox { \epsfxsize=7.cm
     \epsfbox{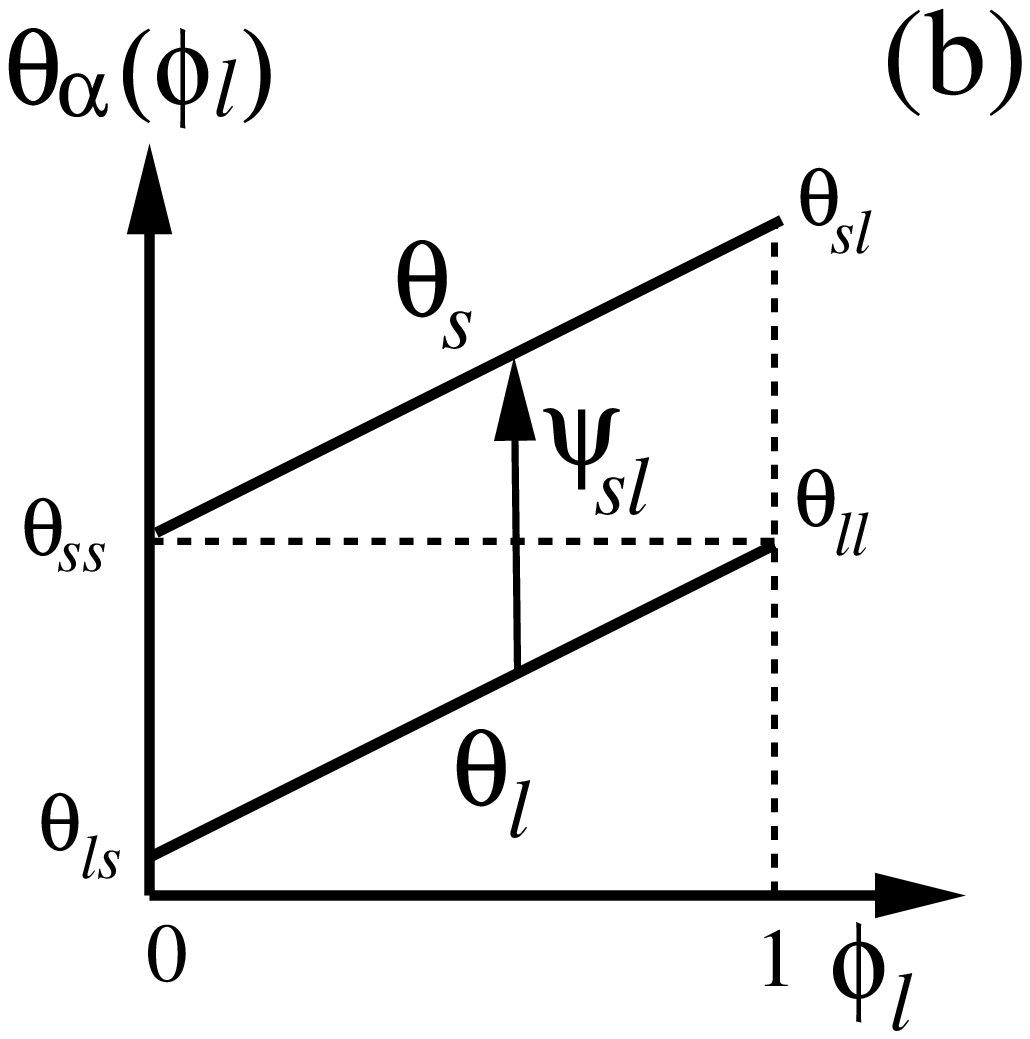} } }
     	    }
\vspace{1cm}
\narrowtext
\caption{Case A: grains differing only in their sizes 
(small type $s$ and large type $l$ grains). 
(a) Diagram of the collision functions. 
(b) Diagram of the generalized angles of repose as a
 function of the volume 
fractions $\phi_l$. The angles of repose $\theta_{ss}$ 
and $\theta_{ll}$ for the 
two pure species are equal. Due to the size difference, 
the cross-angles of repose 
$\theta_{ls}$ and $\theta_{sl}$ are different.}
\label{case1}
\end{figure}

\section{Possible differences between the granular species}
\label{cases}

\subsection{Grains differing only in their size}

Let us consider the case where we mix a species of small grains, denoted
the $'s'$ species, with a species of large grains, denoted the $'l'$
species, both species having similar densities
and the same shape or roughness
(in this case, 
indices  $'1'$ and $'2'$ are replaced by indices $'s'$ and $'l'$). 
The size difference between the two species allows us to compare the
collision functions. 
A grain more easily sets a small grain into motion
 than the reverse, hence
\begin{equation}
\label{ordrex}
 x_s(\theta) < x_l(\theta).
\end{equation} 
A small grain is more easily captured on a surface of large
 grains than the reverse, then 
\begin{equation}
\label{ordrez}
 z_l(\theta) < z_s(\theta).
\end{equation} 
Equations (\ref{ordrex}) and (\ref{ordrez}) imply that the two cross-angles 
$\theta_{\alpha \beta}$, defined by Eq. (\ref{theta12}), satisfy
\begin{equation}
\theta_{ls} < \theta_{sl}.
\end{equation} 
The size difference between the two species does not allow to compare neither 
the functions $a_l(\theta)$ and $a_s(\theta)$, nor the functions $b_l(\theta)$ 
and $b_s(\theta)$.

However considering that the $'l'$ and $'s'$ species have the same surface 
properties, we can obtain other inequalities. 
In first order approximation, 
 the probability that 
an $'s'$ static grain is set into motion by an $'s'$ rolling grain
 is equal to the probability that an $'l'$ static 
grain is set into motion when it is collided by an $'l'$ rolling grain. 
The two probabilities are approximately 
equal because in both collisions the two 
interacting grains have the same weight, and because the two interacting 
surfaces are identical. Hence we have   
\begin{equation}
\label{a=}
a_s \simeq a_l \equiv a.
\end{equation} 
Similarly,
 the probability that a rolling grain is captured by a static grain 
belonging to the same species does not depend on the grain species
being $'s'$ or $'l'$, then
\begin{equation}
\label{b=}
b_s \simeq b_l \equiv b.
\end{equation} 
Equations  (\ref{a=}) and (\ref{b=}) imply that the two pure
angles $\theta_{\alpha \alpha}$ 
(defined by Eq. (\ref{theta11})) satisfy 
\begin{equation}
\theta_{ll} = \theta_{ss}.
\end{equation} 

Thus, the present model shows that two granular 
species with identical surface 
properties have also identical angles of repose, as observed in experiments. 
Finally, the size difference between the two species involved 
in collisions yields
\begin{equation}
\label{ordrea_b}
x_s < a < x_l,  \hspace{0.5cm} z_l < b < z_s.
\end{equation} 
We consider the simplest case by using the  set of simplest relations
consistent with Eqs. (\ref{ordrea_b})
\begin{equation}
\label{ab=mean}
a = (x_s + x_l)/2, \hspace{0.5cm} b = (z_s + z_l)/2.
\end{equation} 
Equations (\ref{ordrea_b}) imply the following relations for the angles of 
repose, as proposed by MCS~[\onlinecite{mcs}]
\begin{equation}
\theta_{ls} < \theta_{ll} = \theta_{ss} < \theta_{sl}.
\end{equation} 

\begin{figure}
      \centerline {  \vbox{  \hbox
      { \epsfxsize=7.cm
      \epsfbox{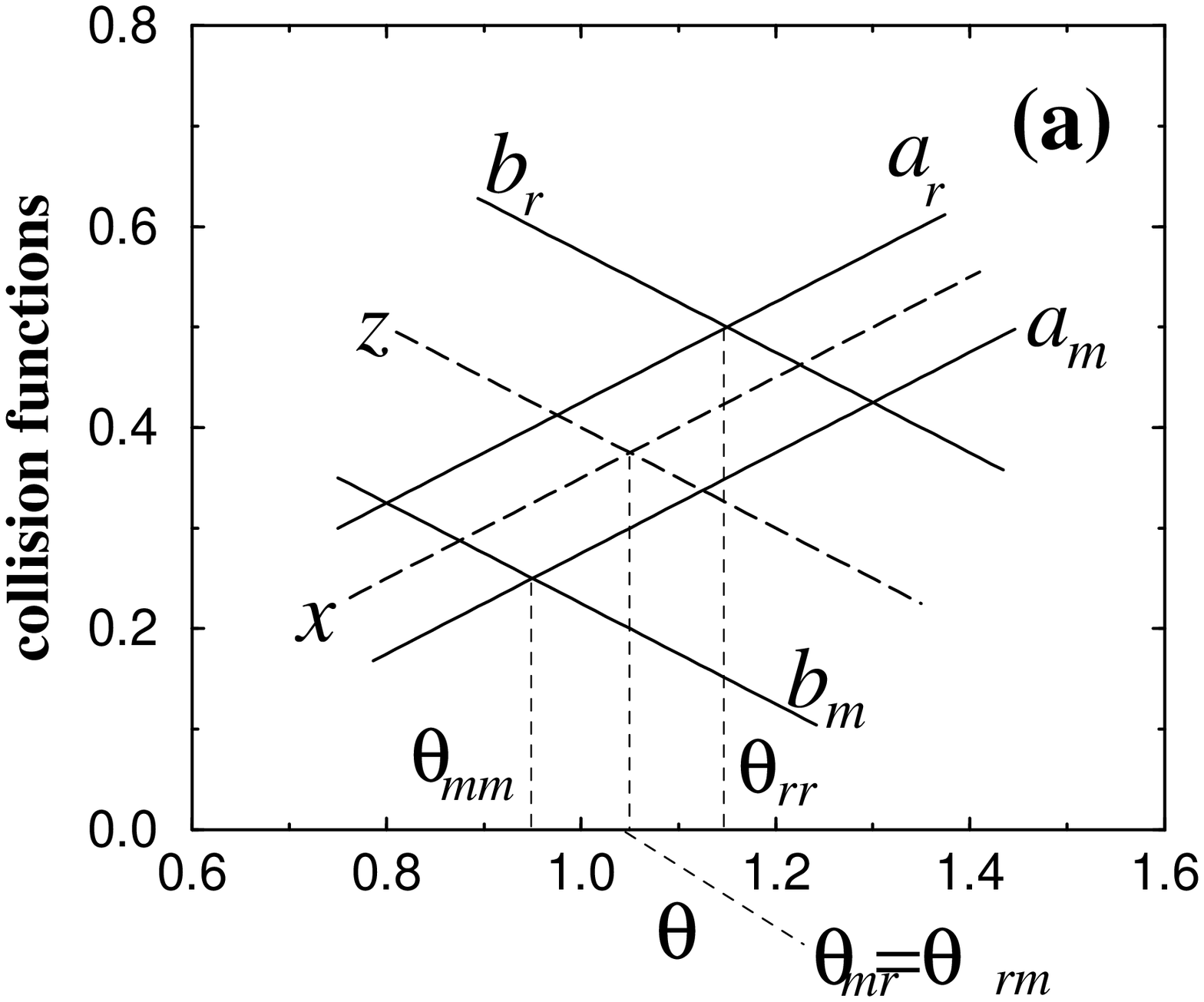} } 
      \hbox
      {\epsfxsize=7.cm
      \epsfbox{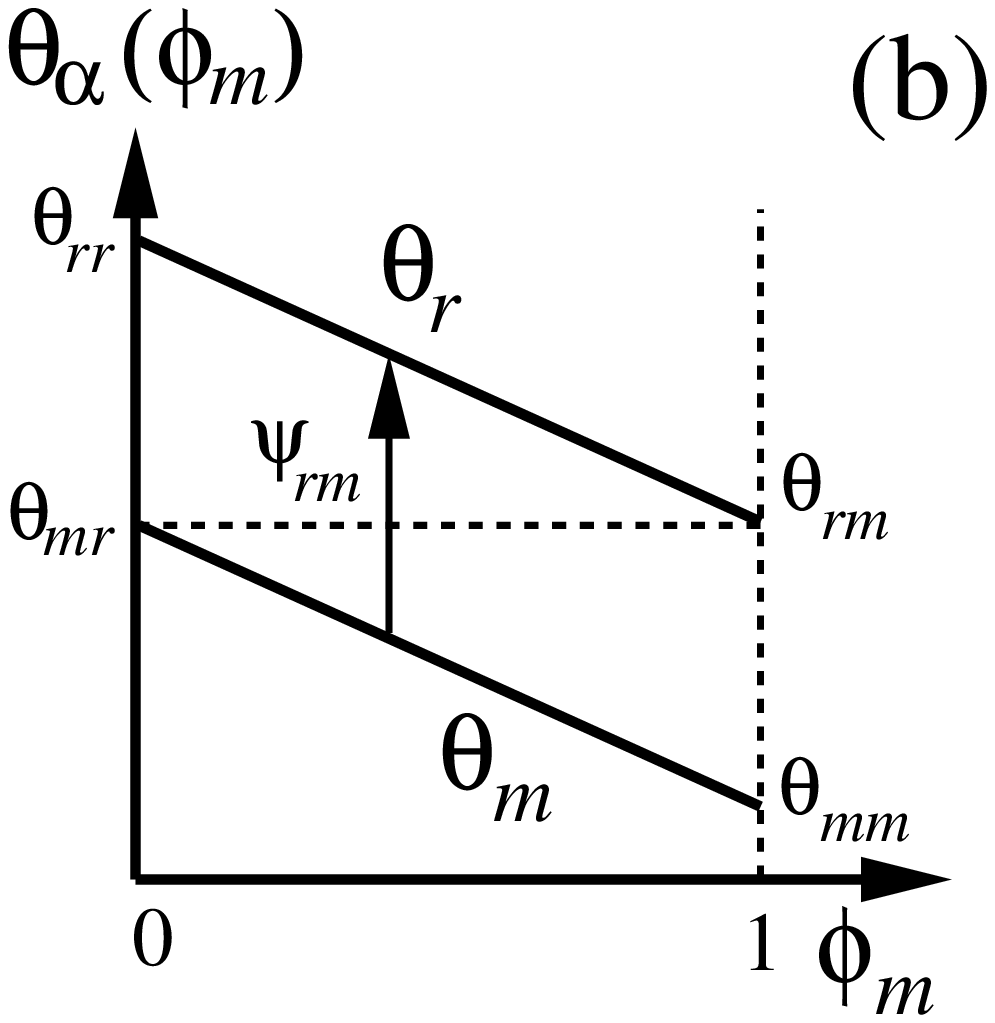} } }
     	    }
\vspace{1cm}
\narrowtext
\caption{Case B: grains differing only in their surface properties 
(rough type $r$  and smooth type $m$ grains). 
(a) Collision functions.  
(b) Generalized angles of repose: due to the surface 
differences the two angles 
of repose $\theta_{ss}$ and $\theta_{ll}$ of the pure species 
are now different, but the cross-angles of repose are the same due to the
equal size of the grains.}
\label{case2}
\end{figure}

The collision functions and the generalized angles of repose 
$\theta_l(\phi_s)$ and 
$\theta_s(\phi_l)$ are represented in Figs.~\ref{case1}a and~\ref{case1}b. 
Note that due to our assumptions (\ref{ab=mean}), 
$\psi_{sl} \equiv \theta_s(\phi_l) - \theta_l(\phi_s)$ is a constant 
independent of $\phi_{\alpha}$.
The angles of repose $\theta_{ss}$ and $\theta_{ll}$ of both pure species
are equal due to their equal shapes, 
but due to the size difference between the particles we have 
$\psi_{sl} > 0$ for any value of 
$\phi_{\alpha}$;
the $l$ rolling species amplifies more easily the rolling phase than the $s$ 
rolling species, because large grains are less easily captured, and more
easily set another grain into motion. The larger the size difference
between the two species, the larger the value of 
$\psi_{sl}$ and the larger the strength of the exchange processes. 
In our case, $\psi_{sl}$ is a small constant so that the linear development 
of the collision functions remains valid.

\subsection{Grains differing only in their surfaces properties}
\label{caseb}

Let us now consider the case where we mix  species of rough grains, denoted
the $'r'$ species, with  species of smooth grains, denoted the $'m'$ species.
The rougher the surface of a grain species, the more 
efficient a collision between 
two grains belonging to this species. Hence we have
\begin{equation}
\label{ordreaANDb}
a_m(\theta) < a_r(\theta),  \hspace{0.5cm} b_m(\theta) < b_r(\theta).
\end{equation} 
Experiments show that  granular species with the 
smoother surface have also the 
smaller angle of repose: $\theta_{mm} < \theta_{rr}$~[\onlinecite{makse2}]. 
This inequality is equivalent to the condition 
\begin{equation}
 a_r - a_m <  b_r - b_m.
\end{equation} 
Thus the last inequality is an experimental constraint for our model.

The different surface properties between the two species does not allow to 
compare neither the functions $x_m$ and $x_r$, nor the functions 
$z_m$ and $z_r$.
However, 
if we also assume that the $'r'$ and $'m'$ species have the same size, 
both cross-interactions correspond to collisions where two grains of the 
same mass interact {\it via} a rough surface in contact with a 
smooth surface.  
In first order approximation, the probability of a 
cross-interaction does not 
change if the two grains involved in the collision are switched, hence 
\begin{equation}
\label{xz=}
x_r \simeq x_m \equiv x,  \hspace{0.5cm} z_r \simeq z_m \equiv z.
\end{equation} 
Equations (\ref{xz=}) imply 
\begin{equation}
\theta_{rm} = \theta_{mr}.
\end{equation} 
Finally, the surface differences between the two species 
involved in collisions yield
\begin{equation}
\label{ordrex_z}
a_m < x < a_r,  \hspace{0.5cm} b_m < z < b_r.
\end{equation} 
We take the simplest relations consistent with Eqs. (\ref{ordrex_z})
\begin{equation}
\label{xz=mean}
x = (a_m + a_r)/2,  \hspace{0.5cm} z = (b_m + b_r)/2.
\end{equation} 
Equations (\ref{ordrex_z}) imply the following relations
for the angles
\begin{equation}
\theta_{mm} < \theta_{rm} = \theta_{mr} < \theta_{rr}.
\end{equation} 
The collision functions 
and the generalized angles of repose $\theta_r(\phi_m)$ and 
$\theta_m(\phi_r)$ are shown in Figs.~\ref{case2}a and~\ref{case2}b.
Here again $\psi_{rm}$ is a small 
constant due to our assumptions (\ref{xz=mean}) and to the validity
of the linear approximations.
Since $m$ rolling grains are less easily captured, they amplify more easily 
the rolling phase than the $r$ rolling grains.

\begin{figure}
      \centerline {  \vbox{  \hbox
      {\epsfxsize=7.cm
      \epsfbox{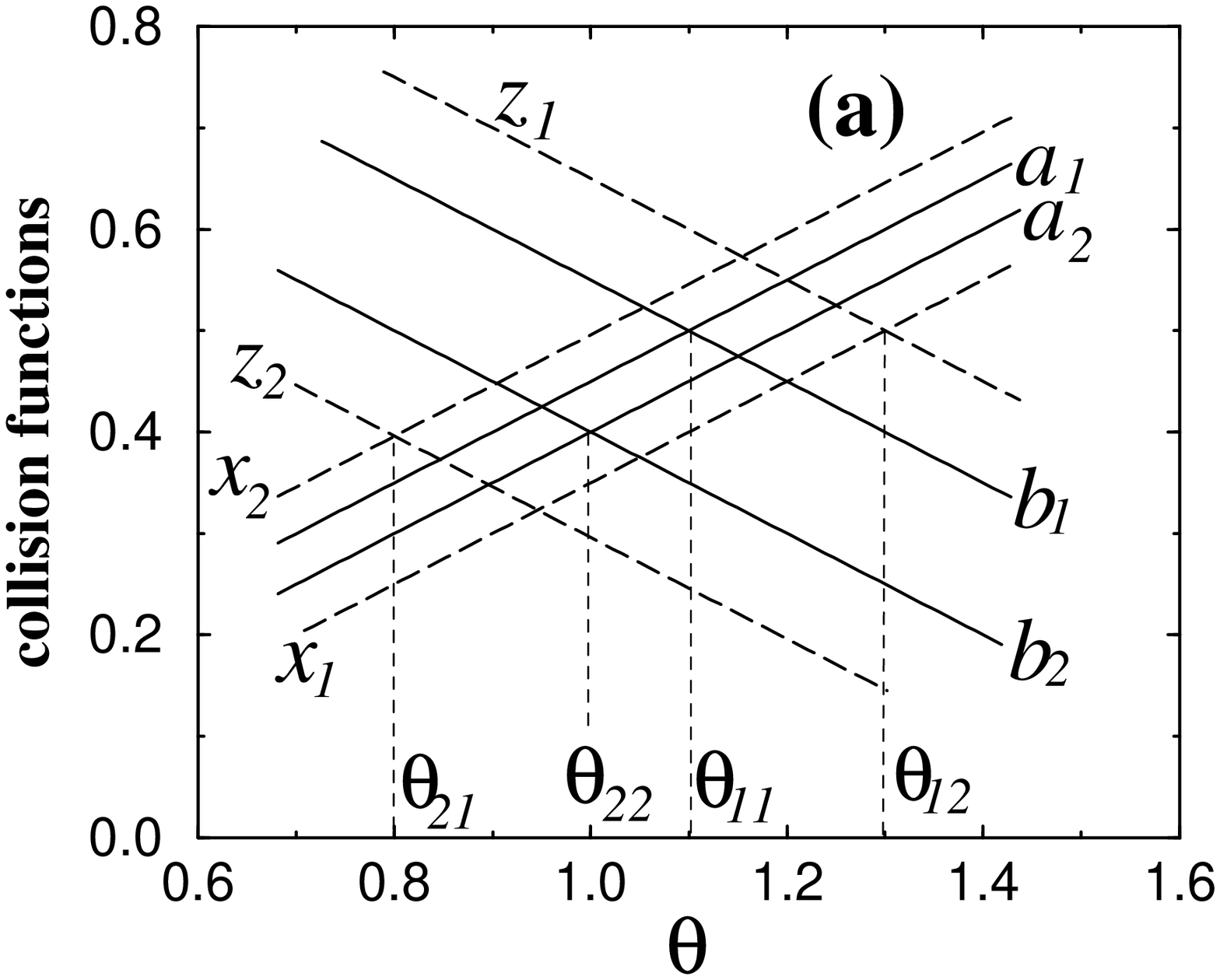} } 
    \hbox
   {\epsfxsize=7.cm
   \epsfbox{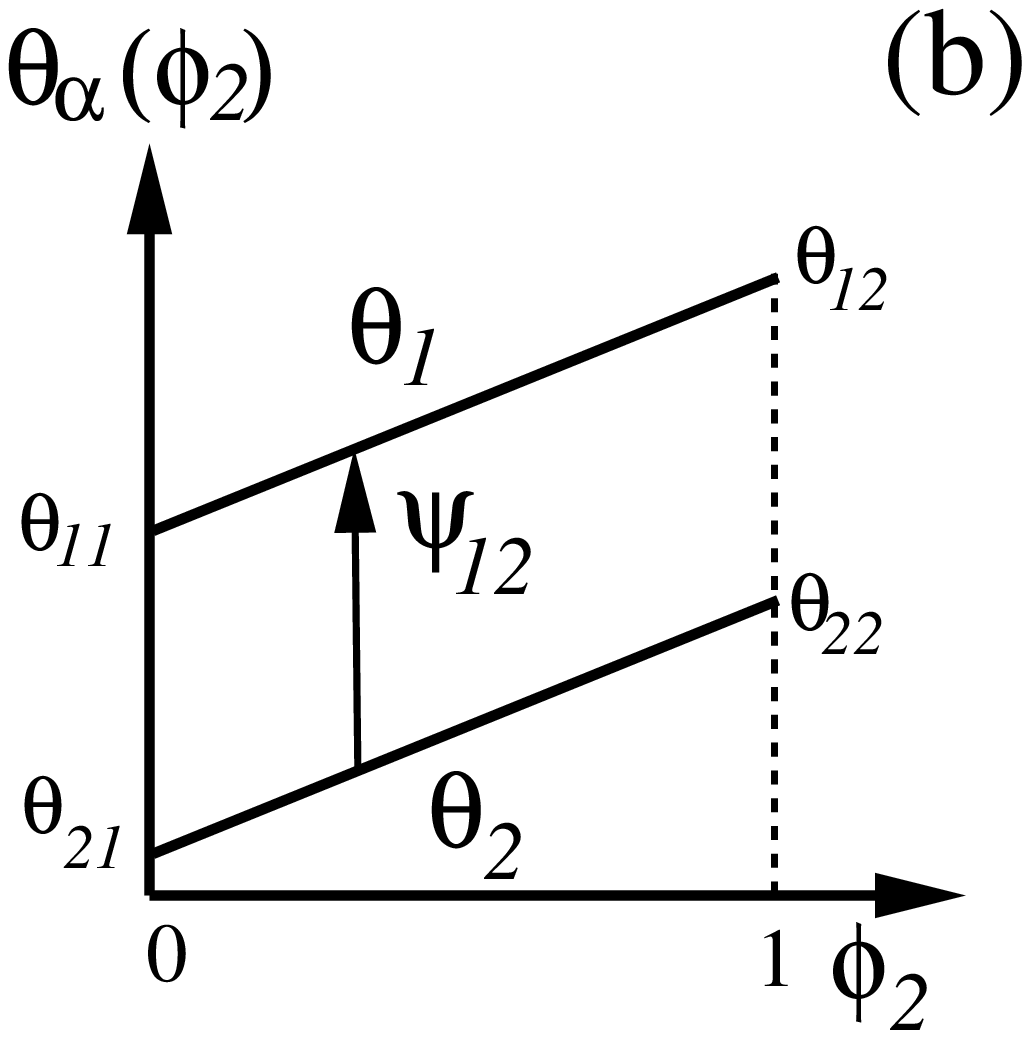} } }
      	    }
\vspace{1cm}
\narrowtext
\caption{Case C: mixture of small rough grains (type $1$)
 and large smooth grains (type $2$). 
(a) Collision functions: when the grains differ both in size and in shape, the 
eight functions are distinct.
(b) Generalized angles of repose.}
\label{case3}
\end{figure}

\subsection{Mixture of small rough grains and large smooth grains}
\label{casec}

We denote ``type $1$'' grains the small rough grains, and ``type $2$'' 
the large smooth grains (here we have $1=s=r$, and $2=l=m$).
The size difference between the two species still yields Eqs.~(\ref{ordrex}) 
and~(\ref{ordrez}), and the cross-angles of repose verify
$\theta_{21} < \theta_{12}$. Moreover, the surface differences still imply 
Eqs.~(\ref{ordreaANDb}), and the angles of repose verify 
$\theta_{22} < \theta_{11}$. 
If the size difference is very small, effects due to the surface differences 
are stronger, and we get $\theta_{22}<\theta_{21}<\theta_{12}<\theta_{11}$. 
This case is close to the situation where grains differ only in their surface 
properties.

However, when the size difference is more important, we obtain a case close 
to the situation where the particles differ only in their sizes. Then, we 
have (as postulated in ~[\onlinecite{mcs}])
\begin{equation}
\theta_{21}<\theta_{22}<\theta_{11}<\theta_{12}, 
\end{equation}
and 
\begin{equation}
x_1 < a_2 < a_1 < x_2,  \hspace{0.5cm}  z_2 < b_2 < b_1 < z_1.
\end{equation}
The collision functions and the generalized angles of repose for this case 
are represented in Figs.~\ref{case3}a and~\ref{case3}b.

\begin{figure}
      \centerline {  \vbox{ \hbox
      {\epsfxsize=7.cm
      \epsfbox{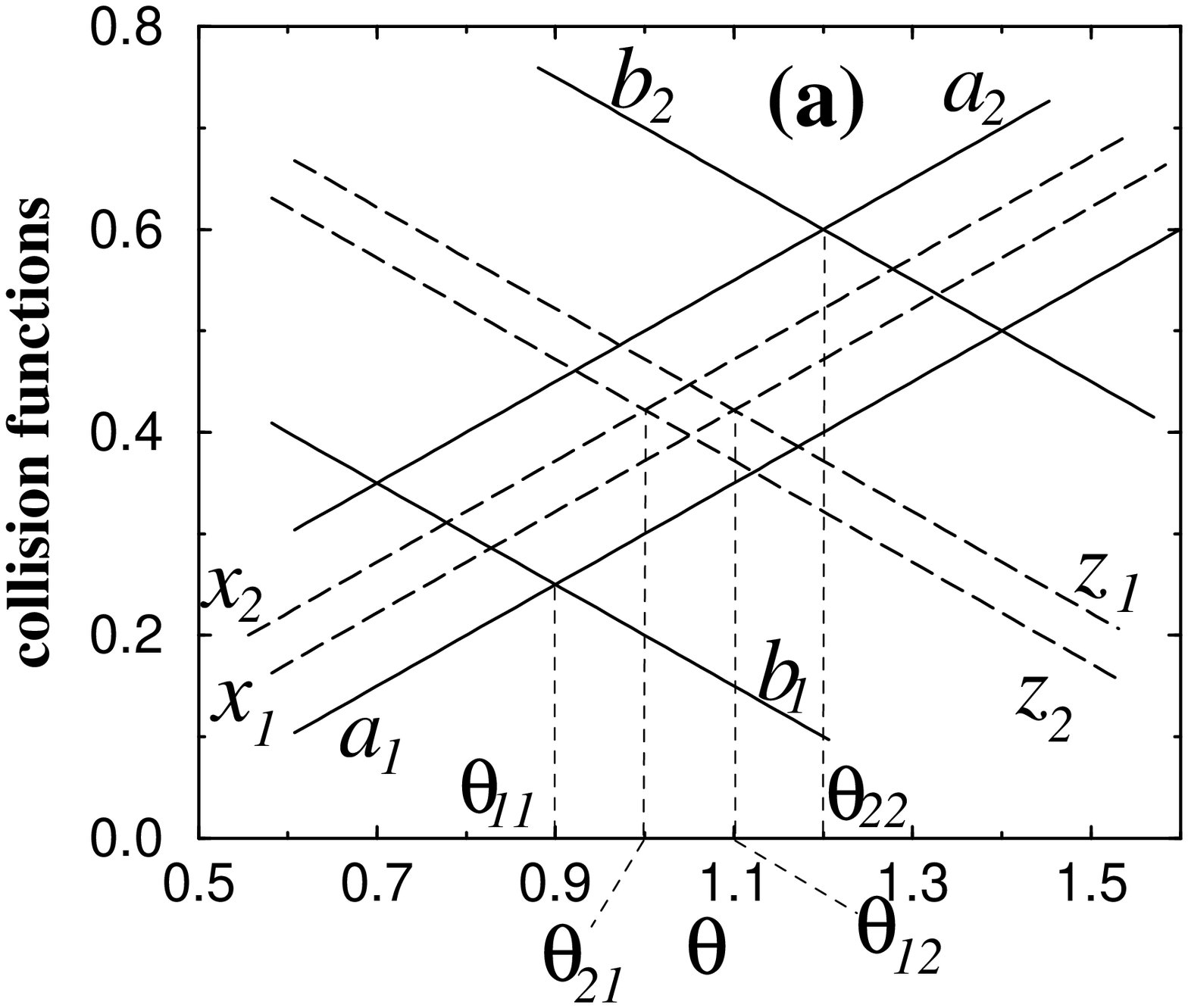} } 
   \hbox
   {\epsfxsize=7.cm
   \epsfbox{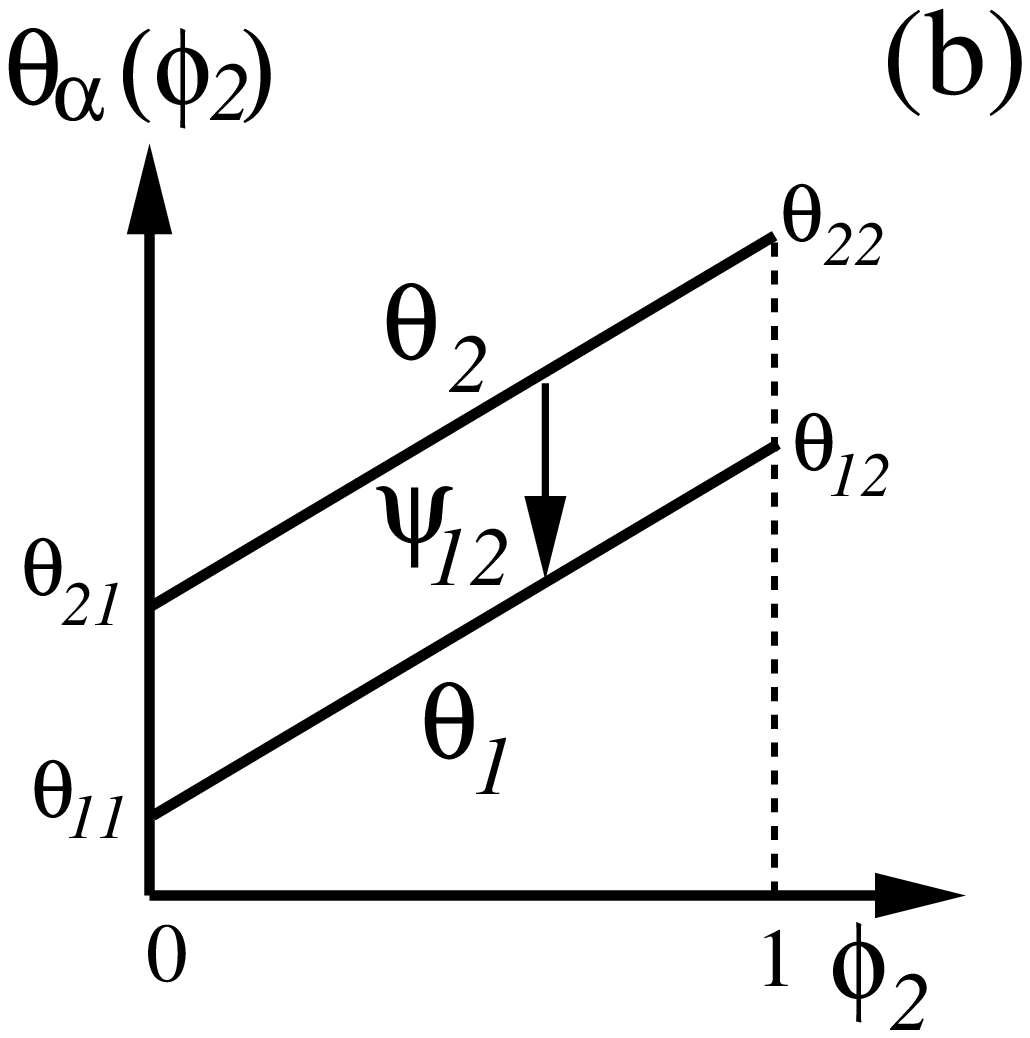} } }
      	    }
\vspace{1cm}
\narrowtext
\caption{Case D: mixture of 
small smooth grains (type $1$), and  large rough grains (type $2$) 
when the size difference is small. 
(a) Collision functions. 
(b) Generalized angles of repose.}
\label{case4}
\end{figure}

\subsection{Mixture of small smooth grains and large rough grains}

We denote ``type $1$'' grains the small smooth grains, and ``type $2$'' 
the large rough grains ($1=s=m$ and $2=l=r$).
As explained previously, we have for the different angles of repose
\begin{equation}
\theta_{11}<\theta_{22},  \hspace{0.5cm}  \theta_{21}<\theta_{12}.
\end{equation}
We consider here the case when the difference in size is small, and shape 
segregation effect is more important than size segregation effect.
Then in this situation, we have (see Fig.~\ref{case4}b)
\begin{equation}
\theta_{11} < \theta_{21}  <\theta_{12} < \theta_{22}.
\end{equation}
As shown in Fig.~\ref{case4}a, the collision functions satisfy
\begin{equation}
a_1 < x_1 < x_2 < a_2,  \hspace{0.5cm} b_1 < z_2 < z_1 < b_2 .
\end{equation}

\section{Segregation in the filling of a silo}
\label{steady}

\subsection{General equations}

\begin{figure}
      \centerline { \vbox{ \hbox
      {\epsfxsize=6.cm \epsfbox{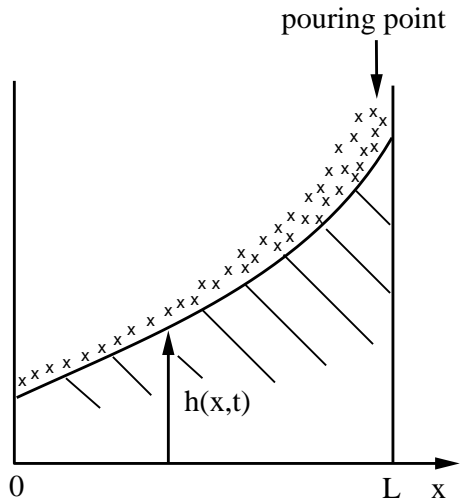} } }
      }
\vspace{1cm}
\narrowtext
\caption{Filling of a two-dimensional silo, in which a 
constant flux of grains is 
poured. In the steady state, the surface of the static grains rises uniformly, 
at the constant speed $w$.}
\label{coordinates}
\end{figure}

In order to describe the segregation between the two species predicted by
the present model, 
we study the case of the steady state filling of a two-dimensional silo,
which is a simple geometry and 
has been the focus of experimental studies 
by different authors~[\onlinecite{makse1,makse2,kaka2,yan}].
Inside a two-dimensional cell made of two vertical plates
separated by approximately 0.5 cm, a mixture of two different species
is poured with a constant in-going flux. Let us call the horizontal axis 
the $x$ axis. The cell is located between $0<x<L$,
and the grains are poured at the point of injection at $x=L$ 
(Fig.~\ref{coordinates}).
We focus on the steady state filling; 
as the mixture is poured in the cell, the surface of the pile 
rises uniformly without deforming, at a constant speed $w$ 
\begin{equation}
\frac{\partial h}{\partial t} = w.
\end{equation}

We review some results  found for this situation 
with the minimal model~[\onlinecite{bdg}]. In the steady state filling,
we have $\dot{R}_\alpha = 0$. Then
equations (\ref{dot_h}) and ($\ref{dot_R}$), and the boundary condition
$R(x=0) = 0$, imply that the total 
thickness $R(x)$ of the rolling phase decreases linearly with respect to the
distance from the pouring point 
\begin{equation}
R(x) = \frac{w}{v}x.
\end{equation}

The total exchange between the two phases 
$\dot{R}_1|_{coll} + \dot{R}_2|_{coll}$ can be 
calculated as a function of the generalized angles of repose, using  
the collision matrix $(\ref{M})$. Then Eq. $(\ref{dot_R})$ yields
\begin{equation}
\label{w=}
- w = \gamma [\theta - \theta_1(\phi_2)] R_1  
    + \gamma [\theta - \theta_2(\phi_1)] R_2.
\end{equation}
In this expression, the order of magnitude of the right hand side is 
$\gamma \psi R \sim \psi w x / d$, where $\gamma \simeq v/d$ [Eq. 
(\ref{gamma})]. 
The ratio of the right
hand side of (\ref{w=}) divided by its left hand side is 
$\sim \psi x / d$. Thus, when $x \gg d / \psi$, the left hand side  can
be neglected, and we get for the local slope $\theta$
\begin{equation}
\label{theta}
\theta = \theta_1 (\phi_2) \frac{R_1}{R} + \theta_2 (\phi_2) \frac{R_2}{R}.
\end{equation}
In particular, in the steady state, the value of $\theta$ lies
between the two generalized angles of repose $\theta_1$ and $\theta_2$. 
By combining Eqs. $(\ref{phi})$,
$(\ref{M})$ and $(\ref{theta})$, we obtain the following expressions of the
volume fractions $\phi_{\alpha}$ in the static phase
\begin{eqnarray}
\label{phii}
\phi_1 &=& \left( 1 + \xi_{12} \frac{R_2}{x_1 R_1 + x_2 R_2} \right) 
\frac{R_1}{R}, \nonumber \\
\phi_2 &=& \left( 1 - \xi_{12} \frac{R_1}{x_1 R_1 + x_2 R_2} \right) 
\frac{R_2}{R},
\end{eqnarray}
where
\begin{equation}
\xi_{12} \equiv \gamma \psi_{12} + \Delta x_{12}
\end{equation}
is a constant in our model. 
We notice that the minimal model of [\onlinecite{bdg}] consists of 
taking the cross-amplification functions constant independent on the angle,
$x_1(\theta)=x_2(\theta)\equiv m$, and $\Delta x_{12}=0$.
With these approximations the minimal model can be solved to obtain a closed
form for the concentrations and the rolling grain profiles.

Equations (\ref{phii}) show that segregation happens
when rolling grains stop on the static phase;  
the volume fraction $R_{\alpha} / R$ of the $\alpha$ grains in the 
rolling phase is different from the volume fraction $\phi_{\alpha}$ 
of the same species in the static phase. Since the quantity 
$R_{\alpha} / (x_1 R_1 + x_2 R_2)$ is always positive, 
if $\xi_{12} > 0$ we obtain $\phi_1 > R_1/R$ and 
$\phi_2 < R_2/R$;   type 1 grains are captured more easily 
than type 2 grains. Conversely,  if $\xi_{12} < 0$ 
we obtain $\phi_1 < R_1/R$ and $\phi_2 > R_2/R$;  
type 2 grains stop more easily. This behavior induces 
segregation everywhere in the static phase; the larger 
$\mid \xi_{12} \mid$ the stronger the segregation.

For instance, let us consider the case where we pour 
an equal volume mixture,  $R_1(x=L) = R_2(x=L)$. 
If $\xi_{12} > 0$, we 
obtain at the top of the slope $\phi_1(x=L) > \phi_2(x=L)$. 
Type 1 rolling grains stop more easily, and the fraction of these grains 
decrease in 
the rolling phase $(R_1 < R_2)$, while type 2 rolling grains will 
preferentially stop 
at the bottom of the slope $\phi_1(x=0) < \phi_2(x=0)$. 
Thus, in this example, the static phase contains mostly type 1 grains 
in its upper part, and type 2 grains in its lower part.

In the lower part of the pile ($x \ll L$), it is possible to quantify more 
precisely the segregation for any kind of poured mixture. 
By doing a linear development 
of the equations, we can look for a power law behavior 
valid at the bottom of the slope, as suggested by the solution
of the minimal model~[\onlinecite{bdg}]. If, for instance, we assume
$\xi_{12}$ to be positive, we expect that $R_1 / R$ tends to zero. We write
for $x \ll L$
\begin{equation}
\frac{R_1(x)}{R(x)} \simeq A x^{\eta} , 
\end{equation}
where $A$ and $\eta$ are two positive constants. 
Then $\phi_1$ can be calculated in first order
approximation with Eqs. (\ref{phii})
\begin{equation}
\label{fit}
\phi_1(x) \simeq \left(1 + \frac{\xi_{12}}{x_2} \right) A x^{\eta} . 
\end{equation}
When we substitute these expressions for $R_1 / R$ and $\phi_1$ 
in the relation 
$\partial R_1 / \partial x = w \phi_1 / v$ 
(obtained from  Eqs. (\ref{phi}) and (\ref{dot_R})), 
we obtain an equation whose compatibility confirms that the type 1 species 
follows a power-law behavior at the bottom of the slope. This equation 
also gives the value of the exponent $\eta$
\begin{equation}
\eta = \frac{\xi_{12}}{x_2(\theta = \theta_{22})}.
\end{equation}
Note that the exponent $\eta$ depends on the granular species. 
This power law behavior shows that our model predicts continuous segregation; 
the concentrations of species vary slowly in the container, both 
types of grains 
remaining present everywhere. This continuous segregation 
corresponds to what is 
observed in experiments, when the two granular species do 
not have a wide 
difference in size~[\onlinecite{yan}].

\begin{figure}
     \centerline { \vbox{ \hbox
      {\epsfxsize=7.cm \epsfbox{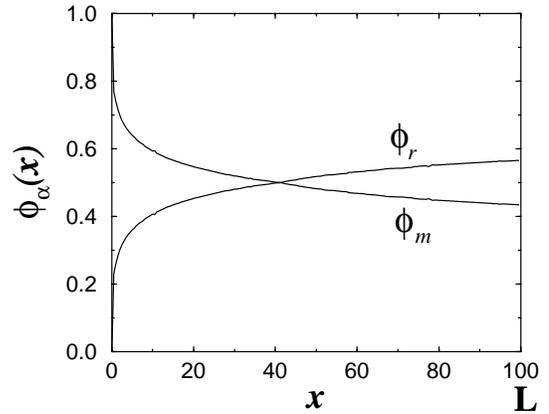} } }
      }
\vspace{1cm}
\narrowtext
\caption{Case B: grains differing only in their surface properties. 
Concentration profiles (calculated numerically) 
during the steady state filling of a silo. 
The segregation clearly appears at $x=L$, where
 $R_m = R_r$ but $\phi_m < \phi_r$.}
\label{SegregationDiffShape}
\end{figure}

\subsection{Segregation for the different mixtures}

For the different mixtures of grains treated in Sec.~\ref{cases}, 
the model predicts continuous segregation during the steady state filling 
of a silo.
In the case of grains differing only in their surface properties, 
we have $x_r = x_m$ and 
\begin{equation}
\xi_{rm} \equiv \gamma \psi_{rm} + \Delta x_{rm} 
              = \gamma \frac{\theta_{rr} - \theta_{mm}}{2} >0 . 
\end{equation}
The sign of $\xi_{rm}$ indicates that inside the static phase, the rough 
grains 
are found preferentially at the top of the silo [see Eq. (\ref{phii})]. 
In order to quantify more precisely the predictions of the model,
we perform a numerical integration of the equations of motion for this case.
Figure~\ref{SegregationDiffShape} shows the results calculated
in the case of an equal volume mixture of rough grains and smooth grains,  
$R_r (x=L) = R_m (x=L)$. For this simulation, we chose $\theta_{rr} = 49^o$, 
$\theta_{mm} = 43^o$, $\psi_{rm} = 3^o$, and $x(\theta = \theta_{rm}) = 
0.375 \gamma$. 
Figure~\ref{SegregationDiffShape} shows
the volume fractions $\phi_{\alpha}(x)$ for $0 \le x \le L$ in the steady 
state.
Note that segregation clearly appears at $x=L$, where $R_m = R_r$ 
but $\phi_m < \phi_r$. Figure~\ref{SegregationDiffShape} confirms that our 
model
predicts a continuous segregation, and not a complete one; 
$\phi_r$ does not fall suddenly  at $x=L/2$, 
but slowly decreases as $x$ decreases. Similar profiles have been recently 
obtained with numerical simulations, by using a granular media lattice gas
model 
to study  the filling process of a two-dimensional silo with inelastic
particles differing in friction coefficients [\onlinecite{antal}].

When the grains differ only in their size, a similar continuous segregation
is found,
with the smaller grains located preferentially at the top of the pile. 
Thus our results indicate that the roughest grains or the smallest ones 
are segregated at the top of the pile, while the largest
grains or the smoothest ones are segregated preferentially at
the bottom. When we mix grains differing both in size and in surface
 properties, 
a competition between size-segregation and shape-segregation appears. 
%

\begin{figure}
      \centerline {  
   \vbox{ 
   \hbox{
   \epsfxsize=7.cm
      \epsfbox{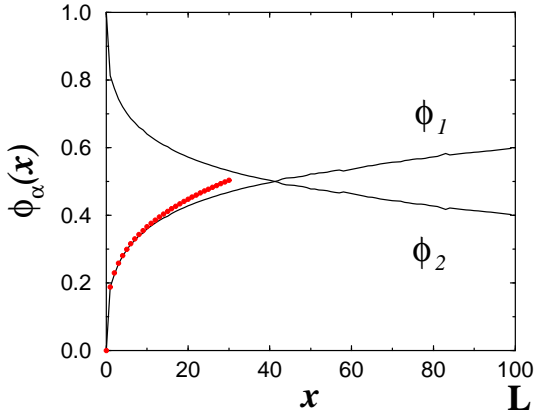} } } }
\vspace{1cm}
\narrowtext
\caption{Case C: small rough grains (type $1$) 
and large smooth grains (type $2$).
Concentration profiles at the surface
calculated numerically in the steady state. The  
dotted line
corresponds to the fit to  the analytical expression (\protect\ref{fit}), 
valid at the bottom of the pile.} 
\label{segregation}
\end{figure}

For the mixture treated in Sec.~\ref{casec} of small rough  grains and 
large smooth  grains, the model predicts continuous segregation  
with the rough and small grains found preferentially at the top of the pile. 
This is the case of strongest segregation found with the present model, 
since both size segregation and shape segregation act simultaneously 
to segregate the small rough grains at the top of the pile
and the large smooth grains at the bottom. 
Figure~\ref{segregation} shows the volume fractions $\phi_\alpha(x)$ 
calculated numerically when an equal volume mixture
of small rough (type 1) grains and large smooth (type 2) grains
is poured in the silo. We set $x_0 = 0.4 \gamma$, 
$\Delta x_{12} = -0.13 \gamma$, $L=100$, and use
the collision functions 
shown in Fig.~\ref{case3}. Figure~\ref{segregation} shows a continuous 
segregation which is 
stronger than in the case of grains differing only in their 
surface properties (case~\ref{caseb}, Fig.~\ref{SegregationDiffShape}). 
At the lower end of the slope ($x \ll L$), we have a complete purification 
of both species due to segregation;  $R_1(x) / R$ and $\phi_1(x)$ tend to zero. 
In that region, we can fit the numerical solution with the 
analytical expression 
Eq. (\ref{fit}) with $\eta = 0.29$.

The last case treated in Sec.~\ref{cases} corresponds to a mixture of small 
smooth grains and large rough grains. The size difference 
is taken to be small, 
to allow for the linear development of the collision functions. Thus
shape segregation is found to dominate over size segregation, so that
the smooth grains are found at the bottom.
However, when the effect of size segregation is comparable to the effect 
of shape segregation, the competition between size and shape segregation 
gives rise to a different phenomenon. Indeed, experiments show that we 
obtain either stratification when the large grains are the roughest, 
or a complete
segregation when the large grains are the smoothest
~[\onlinecite{makse1,makse2,kaka2,yan}]. 
The theoretical model proposed in the present article can explain 
both the complete segregation and the stratification. These phenomena 
are due to the type of
 segregation that  appears directly inside the rolling phase, 
as explained in the next section.

\section{Complete segregation and stratification}
\label{stratification}

So far we have 
treated the cases when the difference between surface properties 
and size of the grains is not too wide, hence
we could perform linear developments 
of the collision functions around the region of interest.
This situation gives rise to the continuous segregation patterns in all
the cases studied in the previous sections.
When the difference between the grains size is important, 
stronger segregation effects are expected.
According to experiments~[\onlinecite{makse1,makse2,kaka2,yan}], this occurs
when $\rho>1.5$, where  $\rho$ is the  ratio of the size
of the large grains divided by the size of the small ones.
The size segregation is due to segregation at the shear surface
between the rolling and the static phase~[\onlinecite{mcs}], 
or it may already happen inside the rolling phase~[\onlinecite{makse2,cms}];  
small rolling grains tend to fall downward through the gaps in between the 
large grains, 
so that they form a sub-layer of small rolling 
grains underneath the large rolling grains. 
This phenomenon is called 
``kinematic sieving'', ``free-surface segregation'' or ``percolation''
~[\onlinecite{segregation4,segregation5,makse2}]. 
The large rolling grains are not in contact with the bulk, and 
are
captured after the small grains.
The percolation effect takes place inside the rolling phase if its 
thickness is larger than (approximately) two or three grain diameters. 
However, even for a thin flow, strong size segregation
occurs in the shear surface between the rolling and static phase
if the size difference between the grains is wide enough.
In both cases, the large rolling grains are captured only after the 
small ones, 
and the collision functions may take the forms proposed in~[\onlinecite{mcs}] 
(see Fig.~4 of~[\onlinecite{m}]).

\begin{figure}
      \centerline {  
   \vbox{ 
   \hbox {\epsfxsize=7.cm \epsfbox{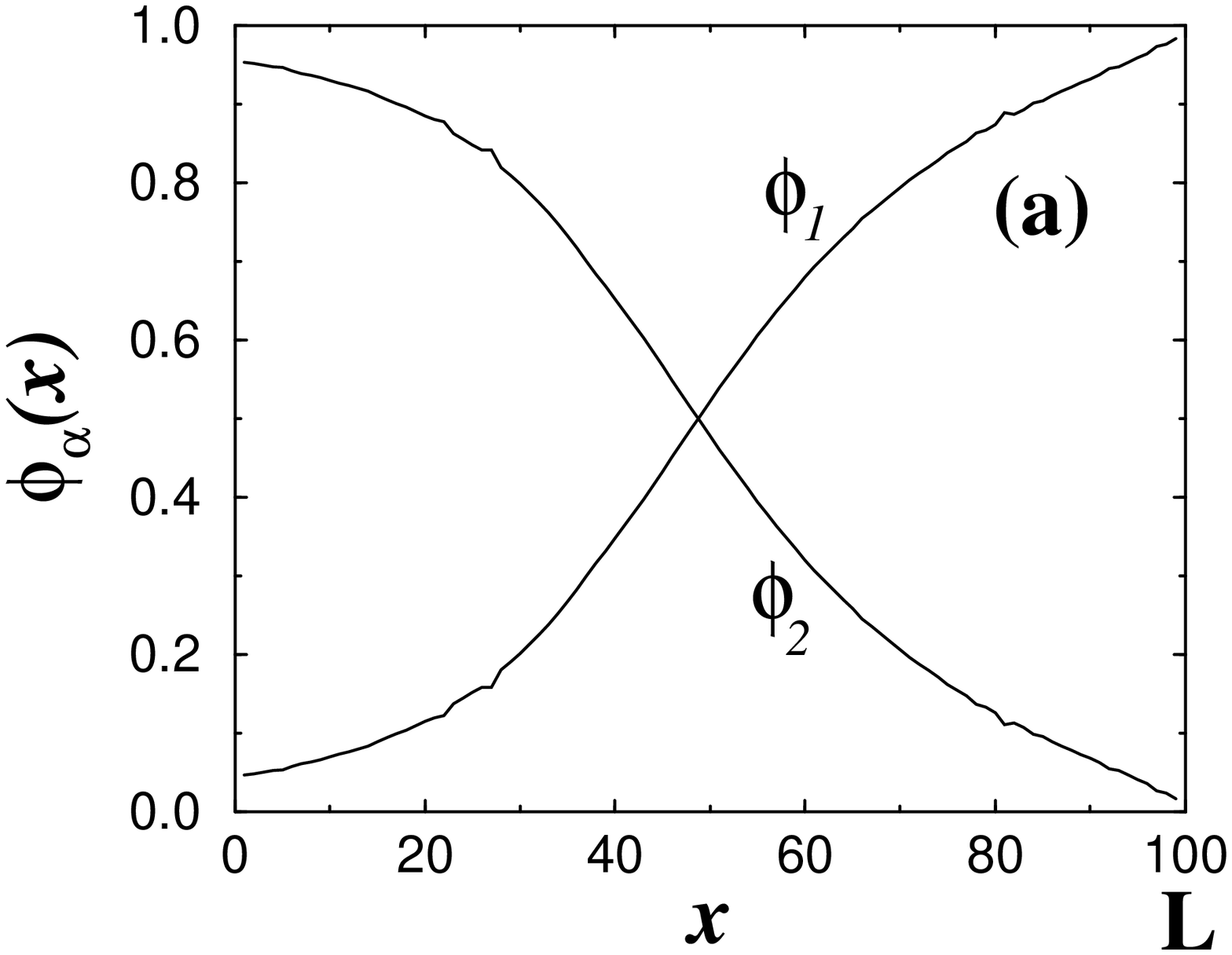} 
   	} 
   \hbox { \epsfxsize=7.cm \epsfbox{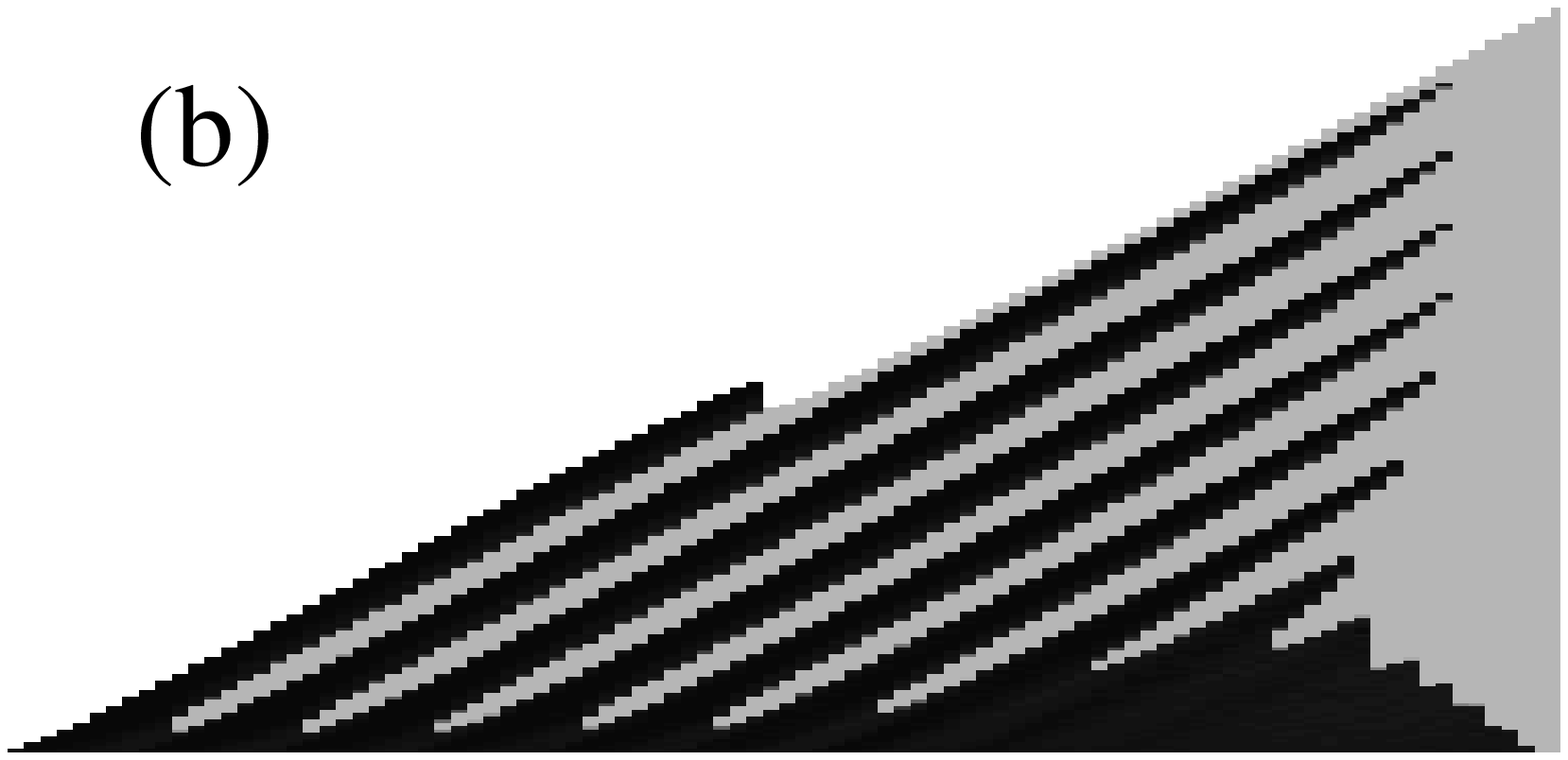} }
   }
   } 
\vspace{1cm}
\narrowtext
\caption{(a) Concentration profiles obtained numerically
showing the 
complete segregation of a mixture of small rough grains (type $1$)
and large 
smooth grains (type $2$)
when the percolation effect is present. 
(b) Stratification of a mixture of large rough grains (dark) and small smooth 
grains (grey), 
when the percolation effect is included in the model.} 
\label{perco}
\end{figure}

Here we use a suitable modification of the interaction term, 
in order to take 
into account the segregation inside the rolling phase within the 
present formalism.
We replace $R_2(x,t)$ in the definition of the interaction term
by  $R_2(x,t) \exp[\lambda R_1(x,t)/R(x,t)]$.
The exponential factor multiplying  $R_2(x,t)$ mimics the fact that 
the interaction of large rolling grains is screened by the presence of small 
grains, so that large 
rolling grains $R_2$ interact with the grains at the surface of the static 
phase only when $R_1(x,t)\ll R(x,t)/\lambda$. The
 dimensionless parameter
$\lambda>0$ measures the degree of percolation
[\onlinecite{cms}].
%
%
We numerically simulate this model using a mixture of grains differing both
in size and shape. When the large grains are smoother than the small grains, 
we obtain a complete  segregation of the mixture  (see Fig.~\ref{perco}a). 
The transition zone between the two species at rest has a size which 
goes approximately 
from a few mm to a few cm. When the large grains are rougher than the 
small grains,
Fig.~\ref{perco}b shows that we find stratification of the static phase.
This stratification results from the competition between size-segregation and 
shape-segregation. 
In conclusion, when our model includes the  percolation effect, 
we are able to reproduce the experimental observations obtained with particles 
of very different sizes~[\onlinecite{makse1,makse2,kaka2,yan}]. 
Our results are also consistent with the ones found by Makse 
{\it et al.}~[\onlinecite{mcs,m,cms}], to whom we refer for further details.

\section{Discussion}

We have proposed an analytical model (called {\it canonical} model) 
that explains the continuous segregation observed during the filling 
of a silo, when the grains have a small difference in size. 
Our predictions could be precisely tested by experimental measurements. 
When the two granular species have a 
wide difference in size, we include, in the 
canonical model, the  percolation effect that appears 
directly in the rolling phase. The model is then able to 
reproduce the complete segregation 
and the stratification observed in experiments; it is also consistent 
with results recently published by Makse {\it et al.} who described 
grains with a wide difference in size.

The canonical model incorporates 
both differences in size and in surface 
properties of the grains, and
 contains several coupling parameters between the two 
granular species such as  the cross-angles of repose, and the two constants 
$x_0$ and $\Delta x_{12}$. In comparison, the minimal
model~[\onlinecite{bdg}] is 
simpler than the canonical model since
 the minimal model 
 contains only one coupling parameter. However,  
as a consequence, the minimal model 
cannot take into account a size difference 
between the particles.

In the present paper, we have applied the canonical model to only one 
practical situation:  the steady state filling of a silo. This situation is 
interesting, since it has already been described experimentally by different 
authors. Moreover, the present model for surface flows of granular mixtures 
could be applied to many different situations. 
In particular, the segregation that appears during thick avalanches 
in rotating cylinders could be studied analytically; the results 
could be compared to the ones published recently for a pure granular 
species~[\onlinecite{TheseTom,BRdG}].

\hspace{1cm}

{\bf ACKNOWLEDGMENTS}

This work has benefited from very stimulating discussions with J.-P. Bouchaud, 
P. Cizeau, Y. Grasselli,  H.J. Herrmann, E. Rapha\"el,  H.E. Stanley,
and S. Tomassone.



\end{multicols} 
\end{document}